\documentstyle[11pt,newpasp,twoside,epsf]{article}
\begin{document}

\title{NUGA: the IRAM Survey of AGN Spiral Hosts}

 \author{S. Garc\'{\i}a-Burillo$^1$, F. Combes$^2$, A. Eckart$^3$, L.~J. 
Tacconi$^4$, 
L.~K. Hunt$^5$, S. Leon$^3$, A.~J. Baker$^4$, P.~P. Englmaier$^4$, F. Boone$^2$, 
E. Schinnerer$^6$, R. Neri$^7$}
 \affil{$^{1}$Observatorio Astron\'omico Nacional (OAN), Madrid, Spain} 
 \affil{$^{2}$Observatoire de Paris, LERMA, Paris, France}
 \affil{$^{3}$Universit\"at zu K\"oln, I. Physikalisches Institut, K\"oln, 
Germany}
 \affil{$^{4}$MPE, Garching, Germany}
 \affil{$^{5}$IRA-CAISMI/CNR, Firenze, Italy}
 \affil{$^{6}$NRAO, Socorro, USA}
 \affil{$^{7}$IRAM-France, Grenoble, France}
\setcounter{page}{111}
\index{Garc\'{\i}a-Burillo, S.}
\index{Combes, F}
\index{A. Eckart}
\index{Tacconi, L.~J.}
\index{Hunt, L.~K.}
\index{Leon, S.}
\index{Baker, A.~J.}
\index{Englmaier, P.~P.}
\index{Boone, F.}
\index{Schinnerer, E.}
\index{Neri, R.}

\begin{abstract}
The NUclei of GAlaxies (NUGA) project is a combined effort to carry out a 
high-resolution ($<$1$\arcsec$) interferometer CO survey of a sample of 
12 nearby AGN spiral hosts, using the IRAM array.  We map
the distribution and dynamics of molecular gas in the inner 1\,kpc of the 
nuclei with resolutions of $\sim 10-50\,{\rm pc}$, and study 
the mechanisms for gas fueling of the different 
low-luminosity AGN. 
First results show evidence for the occurrence of strong 
$m=1$ gas instabilities in Seyferts. NUGA maps allow us to address 
the origin/nature of $m=1$ modes and their link with $m=2$ modes and acoustic 
instabilities, present in other targets.
\end{abstract}
\section{Introduction}
The study of interstellar gas in the nuclei of galaxies is key 
for understanding nuclear activity and circumnuclear star formation.
Within the central kiloparsec, most of the gas is in the molecular phase, which
makes CO lines the optimal tracers of nuclear gas dynamics.  
Up to now, however, CO surveys of galaxies made with single-dish telescopes or 
interferometers were hampered by insufficient spatial resolution. Furthermore, 
until very 
recently (Baker 2000; Jogee et al 2001), the published survey samples have 
included 
very few AGN. 

NUGA aims at filling this gap by carrying out a 
high resolution/sensitivity CO survey of a moderately large sample of 
12 nearby AGN which span the whole sequence of activity types (Seyf 1, 2 and 
LINERs). 
We seek to determine the molecular gas distributions and kinematics in 
the central 1\,kpc of the nuclei with resolutions $<$1$\arcsec$ 
($<$10-50$\,{\rm pc}$), and to study the different mechanisms driving gas 
infall to the AGN. Observations in the 2--1 and 1--0 lines of 
$^{12}$CO are carried out with the IRAM interferometer. On the modeling side, 
N-body simulations are performed to analyse the evolution of 
stellar/gaseous gravitational instabilities in the different study cases. 
NUGA has a multiwavelength approach: the sample is defined based on the 
availability 
of high-quality optical and near-infrared images of the galaxies, both from 
ground-based telescopes and the {\it HST}; these are essential to study how star 
formation
proceeds in the nuclear region. The long term aim is to complete a 
supersample of 25-30 objects observed by consortium members, within and outside 
NUGA, 
with the IRAM array.    

\begin{figure}

\plotfiddle{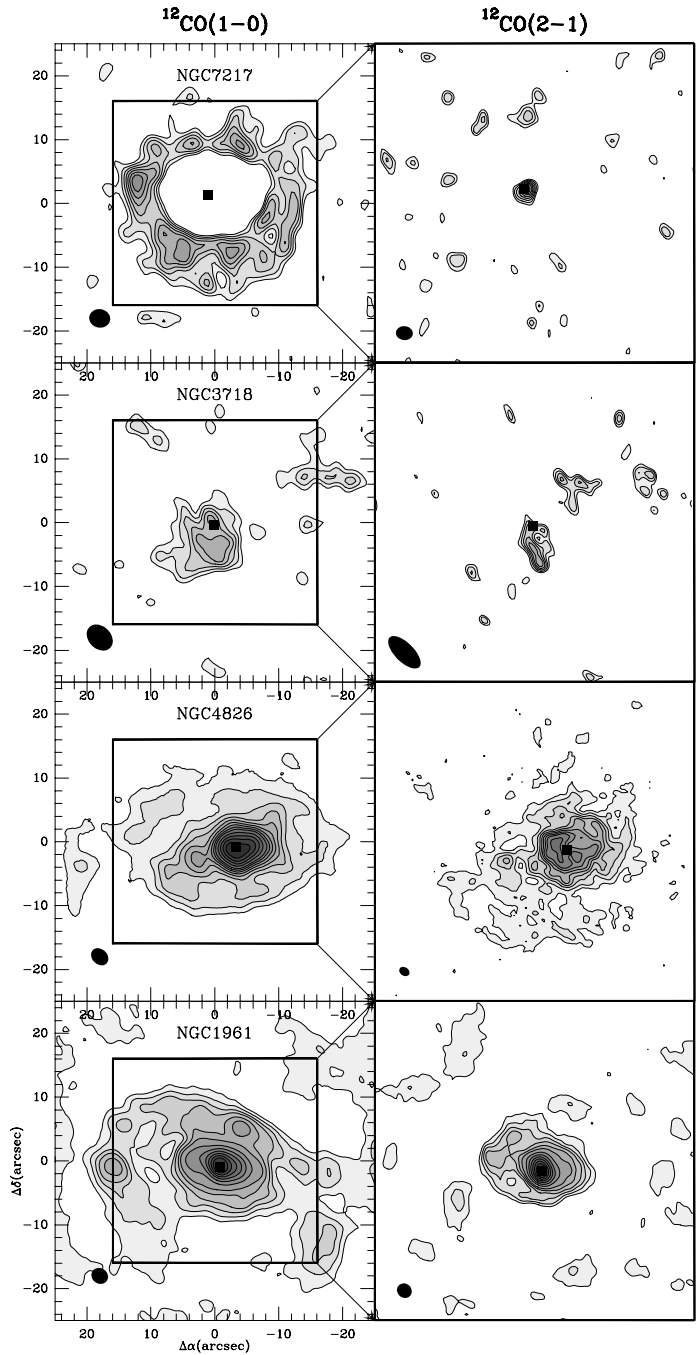}{15.78cm}{0}{65}{65}{-185.7}{-25.2}
\caption{$^{12}$CO(1--0) (left) and $^{12}$CO(2--1) (right) NUGA images, 
discussed in text,
 for NGC\,4826 (Garc\'{\i}a-Burillo et al. 2002), NGC\,7217 (Combes et al. 
2002), 
NGC\,1961 (Baker et al. 2002), and  NGC\,3718 (Krips et al. 2002). CO maps 
unveil a large 
variety of gravitational perturbations: m=1 modes, m=2 modes and stochastic 
patterns.}

\end{figure}

\section{NUGA unveils gravitational instabilities in Seyferts}

Analysis of the first NUGA data on 8 targets has produced a set of studies of 
individual 
galaxies which represent prototypical examples of the large variety of 
gravitational instabilities unveiled in the maps, including m=1 modes (one-arm 
spirals, 
lopsided disks), m=2 perturbations (rings, two-arm spirals) and stochastic 
patterns (non
self-gravitating perturbations) (Fig.1).

\begin{figure}
\plotfiddle{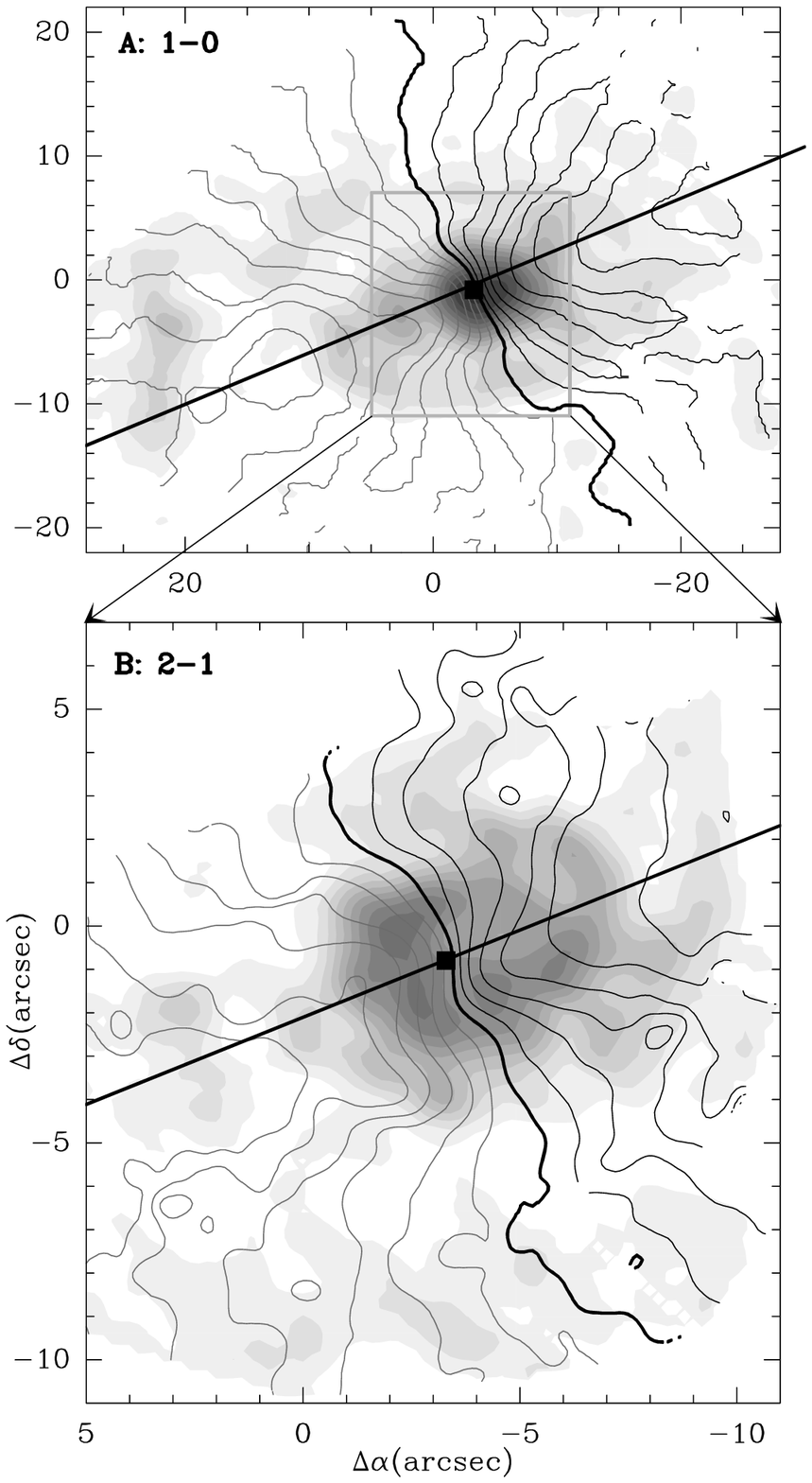}{3.30cm}{0}{40}{40}{-200}{-173}
\plotfiddle{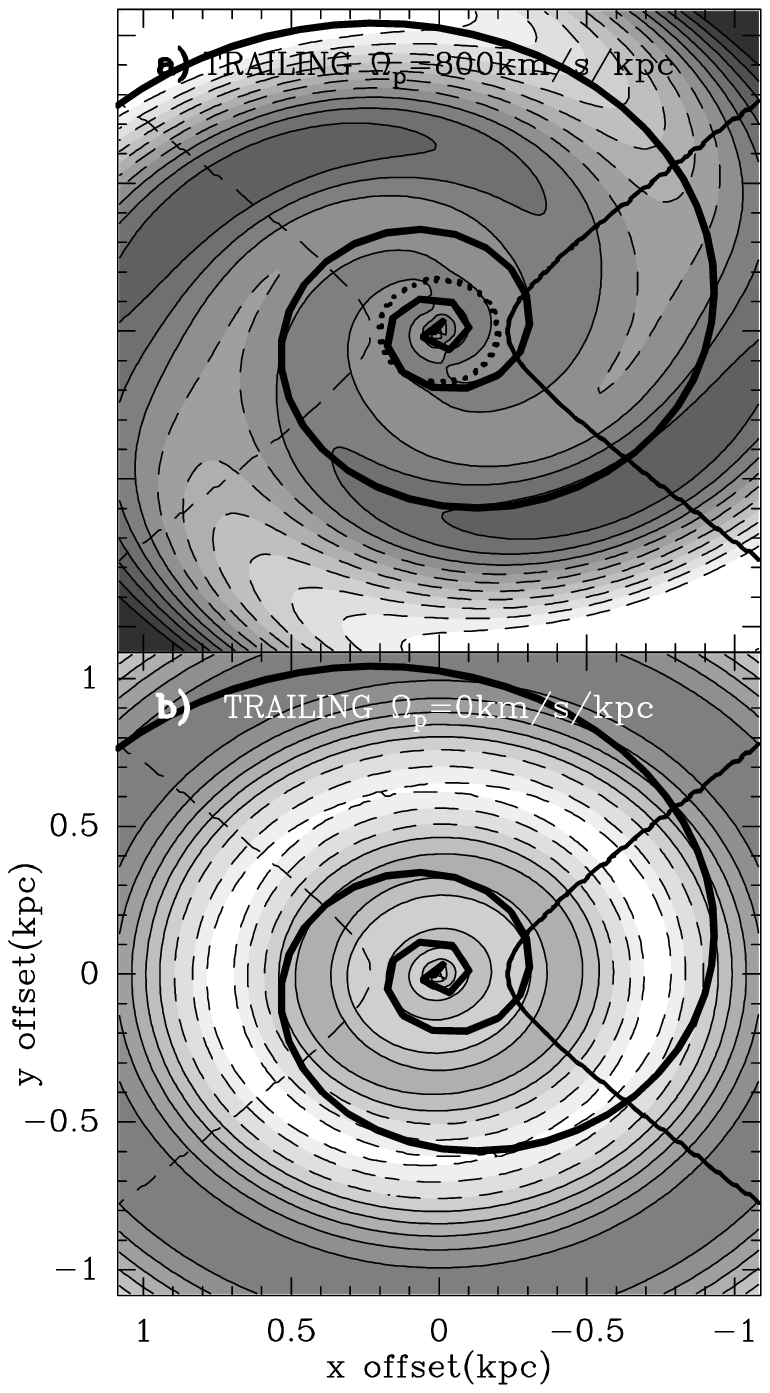}{3.30cm}{0}{57}{57}{-20}{-173}
\caption{Isovelocity CO contours in NGC\,4826 (left pannels) reveal non-circular 
motions. Models (right) allow to identify velocity patterns of $m=1$ modes
 (see Garc\'{\i}a-Burillo et al. 2002).}
\end{figure}

The CO maps obtained in 4 galaxies 
(NGC\,4826, NGC\,1961, NGC\,3718 and NGC\,4579) reveal $m=1$ 
perturbations, which appear as one-arm spirals and lopsided disks.
Asymmetric modes develop from several tens to several hundreds of pc. 
The maps of NGC\,4826 (see Garc\'{\i}a-Burillo et al. 2002) 
reach so far 16\,pc resolution and resolve the inner molecular 
disk to a radius of 700\,pc.  Molecular gas mass is 
distributed in a 40\,pc-radius lopsided disk containing 15\% of the total gas 
content, and two $m=1$ spirals at different radii. A model of the 
streaming motions associated with the $m=1$ perturbations suggest that 
the inner modes  are fast trailing waves (Fig. 2). This 
implies that the lopsided instability may slow down or temporarily halt gas 
infall. 
The $\sim 1.4^{\prime\prime}$ CO(2-1) maps of NGC\,1961 (see Baker et al. 
2002) partly resolve the circumnuclear disk, showing a very strong one-arm 
spiral instability from several hundred pc to 1.5\,kpc.  The Unsharp Masked 
HST/NICMOS map shows also a similar feature in the dust map. 

We have also found examples of $m=2$ 
perturbations (two-arm spiral waves and gas bars). NGC\,6951 shows a very 
regular nuclear spiral structure, roughly coincident with the hot-spot 
circumnuclear ring.
NGC\,2782, a hybrid starburst-transition object, 
shows an intricate mixture of instabilities including an outer two-arm spiral, 
a gas bar, and a nuclear mini-spiral. NGC\,3147 shows a nested two-arm spiral 
structure 
and a compact source on the AGN.   

Most of the nuclear perturbations observed in NUGA targets are 
self-gravitating gas instabilities. The CO+HST images of NGC\,7217 identify 
self-gravitating perturbations in the disk and, also,  
stochastic instabilities on the AGN (see Combes et al. 2002).  The 
1--0 map shows molecular gas confined in 
a very regular circular ring of 800\,pc radius. The 2--1 map shows also a 
compact 
unresolved source on the AGN linked to a
stochastic mini-spiral seen in the $V-I$ HST color image. 
Combes et al. (2002) have made self-consistent numerical 
simulations that account for the CO ring morphology and the onset of central 
non self-gravitating perturbations. The combination of 
ring+stochastic waves in NGC\,7217's gas distribution might reflect evolution 
along the Seyfert sequence, with this galaxy being a more evolved Seyfert.   

\section{Conclusions and Prospects} 

Preliminary results based on still incomplete observations of 8 galaxies of the 
sample reveal a wide range of gravitational instabilities in the central 1\,kpc 
of NUGA targets. Most remarkably, $m=1$ modes appear at different spatial scales 
in some AGNs, although these modes might not {\it universally} favor AGN 
feeding.    
Point-symmetric $m=2$ perturbations and stochastic patterns dominate the 
response 
in others. Although general conclusions will have to wait for a global analysis 
of 
the 30-galaxy supersample, these results already indicate 
that the correlation between activity type (Seyf 1, 2 or LINERs) and nuclear 
morphology of the host might be weak at scales of ten-to-hundred pc. If 
confirmed,
the lack of a clear evolutionary scenario may reflect a mismatch between the 
episodic 
AGN duty cycle, and the larger time-scales needed to build up   
nuclear gravitational instabilities which are directly or indirectly involved 
(by 
the onset of nuclear starbursts) in fueling AGN.

\end{document}